\documentclass[twocolumn]{article}

\usepackage{graphicx}
\usepackage{url}

\begin{document}
\title{Honeypot-powered Malware Reverse Engineering}
\date{}

\author{
	Michele Bombardieri\\
	{\small Open Knowledge Technologies s.r.l.}\\
	{\small Piazza Vermicelli}\\
	{\small 87036, Rende (CS) -- Italy}\\
	\url{michele.bombardieri@gmail.com}\and
Salvatore Castan\`o\\
{\small Open Knowledge Technologies s.r.l.}\\
{\small Piazza Vermicelli}\\
{\small 87036, Rende (CS) -- Italy}\\
\url{salvatorecstn@gmail.com}\and
 Fabrizio Curcio\\
{\small Centro di competenza ICT-SUD}\\
{\small Piazza Vermicelli}\\ 
{\small 87036, Rende (CS) -- Italy}\\
\url{ curcio.fabrizio@gmail.com}
%	%% 2nd. author
\and
 Angelo Furfaro\\       
{\small DIMES -- University of Calabria}  \\
{\small 87036, Rende (CS) -- Italy }\\
\url{ a.furfaro@dimes.unical.it}%
\and
Helen D. Karatza\\
{\small Department of Informatics - Aristotle University of Thessaloniki}\\
{\small	54124 Thessaloniki -- Greece}\\
\url{karatza@csd.auth.gr}
}
\maketitle

\begin{abstract}
	
	Honeypots, i.e. networked computer systems specially designed and crafted to mimic the normal operations of other systems while capturing and storing information about the interactions with the world outside, are a crucial technology into the study of cyber threats and attacks that propagate and occur through networks. Among them, high interaction honeypots are considered the most efficient because the attacker (whether automated or not) perceives realistic interactions with the target machine. In the case of automated attacks, propagated by malwares, currently available honeypots alone are not specialized enough to allow the analysis of their behaviors and effects on the target system. The research presented in this paper shows how high interaction honeypots can be enhanced by powering them with specific features that improve the reverse engineering activities needed to effectively analyze captured malicious entities. 
\end{abstract}

\section{Introduction}
\label{sec:intro}

Nowadays, an ever growing number of systems and devices interoperate and cooperate over the Internet which has become a huge and very complex open distributed system.
Being open is one of the features that brought Internet 
to the role it plays into the current IT scenario.  
However, this expose the interconnected services and devices to the risk of being the target of Internet-enabled malicious entities also known as \textit{malwares}. 
A challenging goal of the research efforts directed in the Cyber Security field is the devising of effective techniques and tools  able to face and defeat the attacks coming from such entities. Hence, it is  crucial to 
analyze and understand a malware behavior in order to
identify the vulnerabilities it exploits and to take the suitable countermeasures. 
%Nowadays running a big network constantly under attack gives the security researchers and auditors a huge amount of informations to handle.
Malware activities are often detected a lot of time after the victim system has been violated and then it is very difficult to trace back the actions that have been executed.
%When the researcher work is focused on malware’s actions and behaviors, it becomes even more difficult to understand its actions on the victim system. 
A malware usually modifies file systems, starts processes, initiates network connections and steals information. All this actions need time to be understood starting just from a binary file, which often is the only tangible record of the malicious code and which need to be analyzed once it has recovered from the victim filesystem.

A honeypot~\cite{Spitzner2003} is a decoy system  which aims at looking like a real networked resource to the world outside. It manifests a low or absent protection level so as to draw attackers' attention and to gather information about their behavior while they gain access to and interact with it. Most of the honeypots currently used are classified as having \textit{low interaction} \cite{MairhEtAl2011, Mokube2007}, which means that they host only emulated services. With such a type of honeypots, it is hard to study malware propagation in details because emulating a service behavior is not a simple task and most of these fake services offer only a, somewhat limited, support  to network traffic analysis, command execution tracking and file modification tracing.
Such type of honeypots are the most diffused, however
smartly crafted malwares and skilled attackers are able to recognize them and  avoid of being caught or analyzed \cite{SysmanEtAl2015}.

On the other side, \textit{high interaction} honeypots are based on real operating systems. This factor lets the malware or the attacker manifest his behavior completely without being stopped by the lack of unimplemented features which can be present in low interaction honeypots. High interaction honeypots are clearly more difficult to handle because the attack surface becomes very large along the fact that it is nearly impossible to predict the attacker or malware intentions with no previous knowledge.

We argue that enhancing  high interaction honeypots by adding reverse engineering instrumentation features will result in a great empowering of  malware analysis processes.
This paper illustrates the design and the features  of \textsc{HERESy} (Honeypot Embedded Reverse Engineering System), a modular high interaction honeypot system whose components work at various levels: file system, network and process execution tracing.  \textsc{HERESy} is able to create on-demand  an isolated environment for each attacker or malware that accesses it by purposely exploiting the notion of container \cite{docker}. For each isolated instance, the system tracks all the attack lifecycle. The attacker or the malware tries to connect to a particular service, if the targeted service is handled by our system, then a proxy intercepts the request and builds on the fly a container redirecting all the traffic to this instance. From this point on all the actions performed by the attacker are tracked and saved on the \textsc{HERESy} logging and storage system.

The rest of this paper is organized as follows. Section~\ref{sec:arch} presents the architecture of \textsc{HERESy}, the roles of its components and how they work together. Section~\ref{sec:exp}  shows a practical example of a malware being captured by the system and subsequently analyzed. Section~\ref{sec:conc} draws the conclusions and the future work.

\section{HEReSy Architecture}
\label{sec:arch}

\begin{figure}[tbp]

\includegraphics[width=1.0\columnwidth]{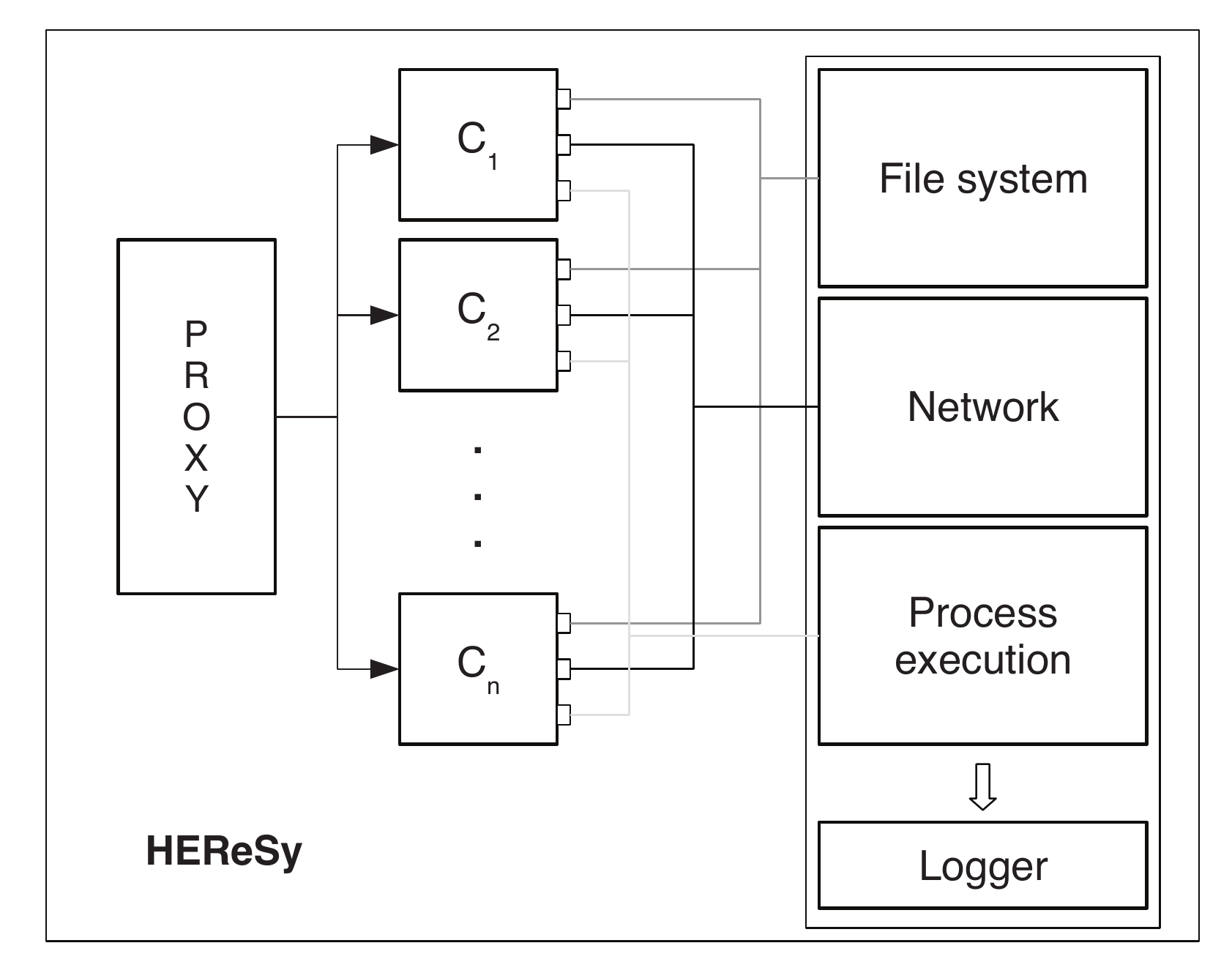}
\caption{\textsc{HERESy} architecture}
\label{fig:harch}
\end{figure}

\begin{figure}[tb]

	\centering
	\includegraphics[width=0.98\columnwidth]{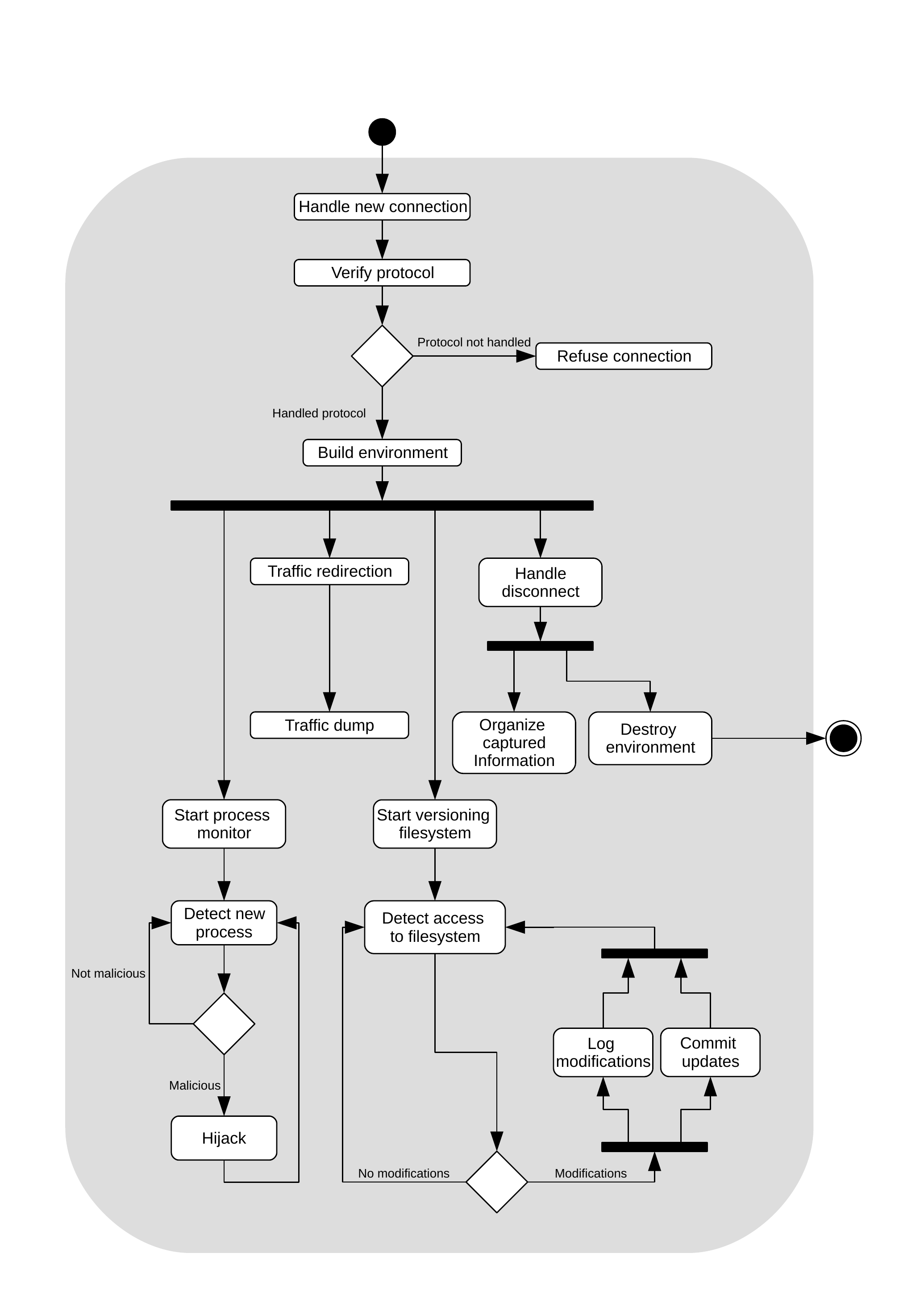}
	\caption{\textsc{HERESy} process}
		\label{fig:wf}
\end{figure}

As pointed out in the previous section, the \textsc{HERESy} architecture, depicted in   the diagram of Fig.~\ref{fig:harch}, is composed by several modules,
%handled by the system. 
each of which  is in charge of  performing a specific hijacking task on the isolated running containers~\cite{docker}.
%into which the attackers log in. 
The \textit{proxy} module plays the role of  container factory: it builds up on-demand a specialized container on the basis of the protocol over which the potentially malicious remote entity tries to access the system and then it redirects all the subsequent incoming traffic to it. At this point, \textsc{HERESy} begins the hijack: file system modifications are tracked by the \textit{filesystem  module} which exploits versioning control technologies; traffic analysis is performed
by the \textit{network sniffing} module; extraneous binaries,  in case they are downloaded by malwares or attackers, are detected by the \textit{process execution} tracing module. Fig.~\ref{fig:wf} depicts an activity diagram that models the \textsc{HERESy} execution workflow which handles connections attempts.

\subsection{Proxy module}
\label{sub:proxy}
Once a connection attempt reaches the system, if the proxy has been configured to handle the requested service, a fresh new container is built on the fly and all the attacker's traffic is redirected to it.
Each attacker will run and operate inside its own isolated environment. When the attacker leaves, the environment is destroyed after all the relevant data have been suitably stored. To speed up the build operation, and hence to improve the \textsc{HERESy} responsiveness, an environment for each service is always up and maintained ready to be used.

%% Cosa aggiungiamo?

\subsection{Filesystem module}
\label{sub:fs}
\begin{figure}[bt]

\centering
	  \includegraphics[width=1.0\columnwidth]{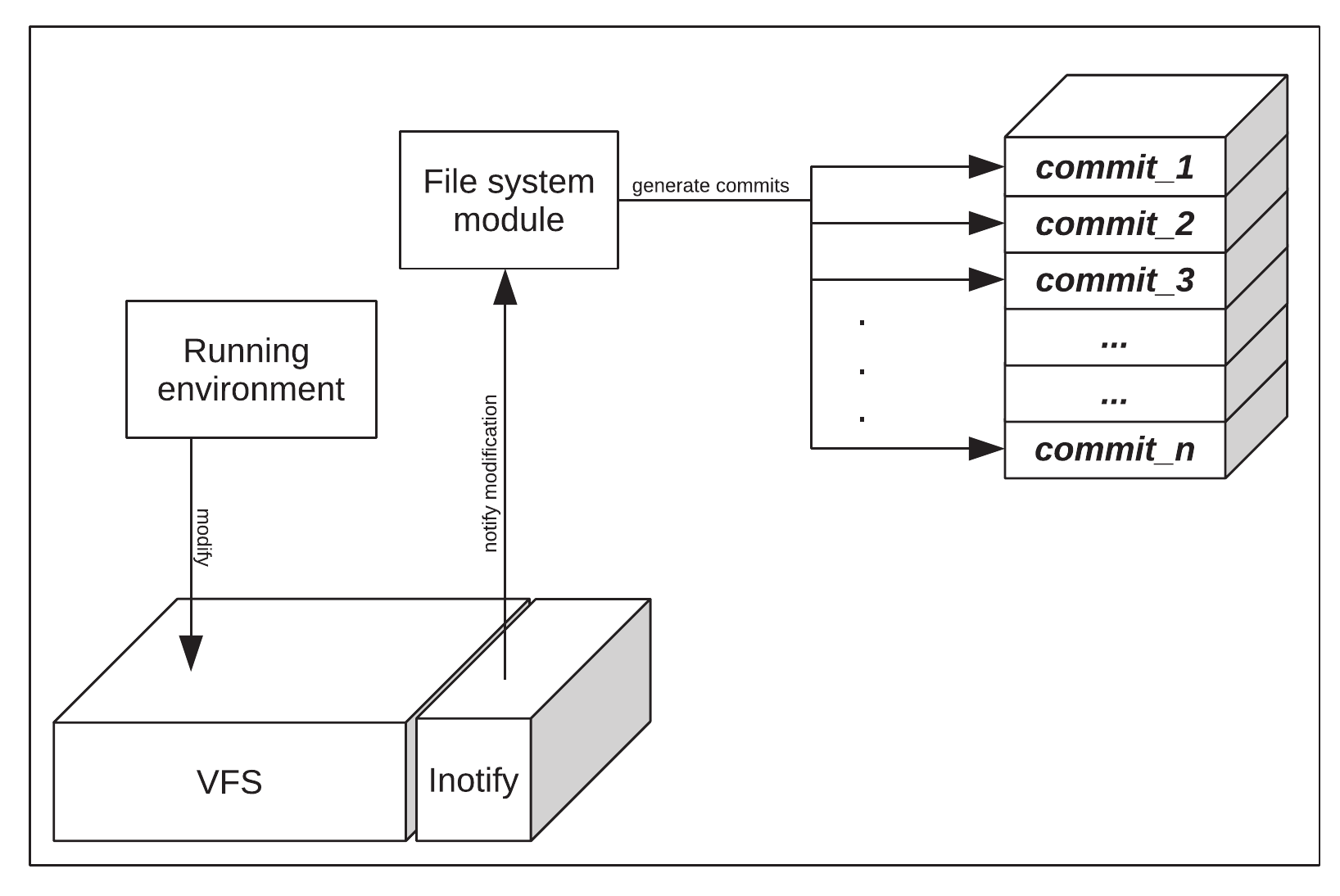}
\caption{ File system module}
\label{fig:fsmod}
\end{figure}

One of the main problem with malicious code is that more than often it modifies the content of file system resources. It is possible to track these modifications by writing custom file systems or by hijacking function library calls. However, the former is an hard task to accomplish because of the low level machinery around the host operating system, the latter can be ineffective because, if the malware calls into the kernel bypassing the libraries,  interception does not occur at all. 
The  solution devised by \textsc{HERESy} consists in exploiting versioning control software on the existing virtual file system layer. File system resources are shared between the host \textsc{HERESy} system and the current attacked environment. By
 using file system notification mechanism \cite{fsnotify}, each time a modification,  creation or deletion happens, the system triggers the versioning control software which tracks the updates.
This technique allows  to  explore the file system history going backward and forward in time. This indeed is crucial for tracing back the malware behavior from the viewpoint of the action made on the file system.  
The file system module of the current version of \textsc{HERESy}  uses of GIT as a versioning control software and the Linux \texttt{inotify} API to handle events \cite{fsnotify} (see Fig.~\ref{fig:fsmod}). Previous experiments used FUSE \cite{fuse} as a file system layer, however the integration with docker was hard to achieve. Another project relaying on GIT in a similar way is \texttt{gitfs}, a FUSE file system which automatically converts all the changes made into commits~\cite{gitfs}.

\subsection{Network module}
\label{sub:net}

Traffic generated by attackers or malicious software is of course fundamental to track where the malicious activities starts and where they try to propagate. For each environment a complete network traffic dump is performed, so a clean data representation of the network flow is given to the threat analyst. The system spawns a daemon for each sandbox environment which is built up. The module relays on the \texttt{libcap} library and saves the attacker generated traffic into the classical \texttt{pcap} format. So using the berkeley packet filtering~\cite{McCanne1993} we are able to select just the traffic generated by the target container.

\subsection{Process execution module}
\label{sub:proc}
For each kind of environment all the binaries are hashed and the signatures are saved into the core system memory. Each time a malicious software or the attacker executes a binary, the command is logged. Also, the module performs the binary hash and if the calculated signature is not present into the white list, the module starts tracking the process using the dynamic binary instrumentation \cite{Hazelwood2011}, so it dumps instructions and memory accesses saving the entire hostile program execution flow.

\begin{figure}[btp]

\includegraphics[width=1.0\columnwidth]{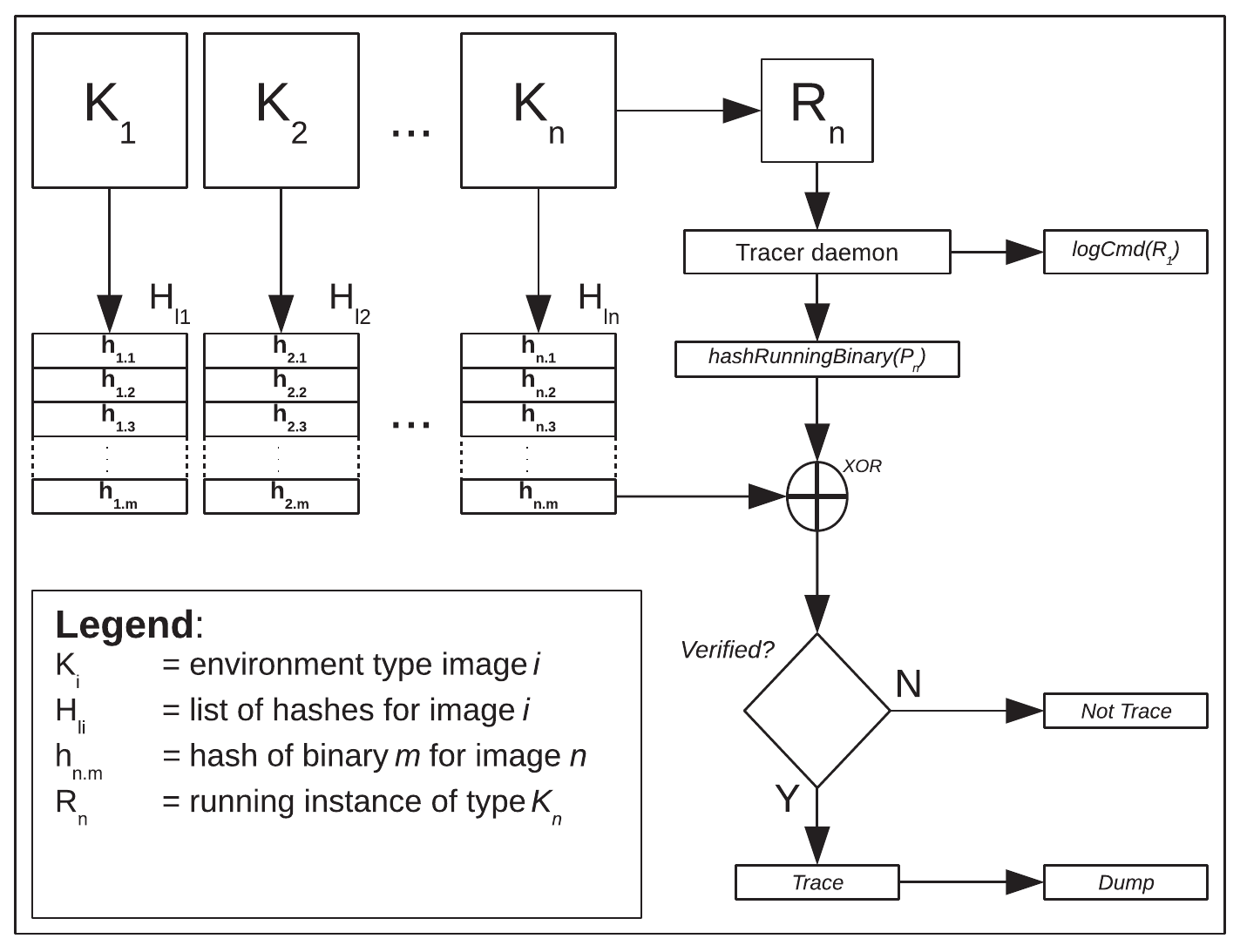}
\caption{Process execution tracing module}
\label{fig:procexec}
\end{figure}
Figure~\ref{fig:procexec}, shows the process execution tracing step. We use for the hashing process SHA512 which is considered strong enough to avoid collisions, so it becomes difficult for an attacker to generate a binary which has the same hash to trick the system. 

In practice, we used both CRIU~\cite{criu2014} and DynamoRIO~\cite{Bruening2003a} in order to trace the process execution and to dump its resources. %Because this two  pieces of software relay both on the \texttt{ptrace} system call is not simple to make they work together.
%once a process is recognized as hostile
The following strategy has been employed to alternate our \texttt{dynamoRIO} based tool with \texttt{CRIU}:  the process is first run inside the \texttt{dynamoRIO} based tool, which detects whether the process is hostile or not on the basis of the signatures of the executable images used by the process; in the case of a not trusted process, the tool begins to dump all the executed instructions and if detects a memory write it asks \texttt{CRIU} to dump all the process resources and waits for its termination; \texttt{CRIU}
receives dump requests by means of a daemon which listen on a socket for messages coming from the \texttt{dynamoRIO} based tool; %
%the process context is saved and a \texttt{pause()} system call is injected. While the hostile process is paused, the tool  forks and starts \texttt{CRIU} which dumps all the process resources and \texttt{wait()} for CRIU to terminate.
once \texttt{CRIU} completes its task, it notifies the \texttt{dynamoRIO} tool which then restarts its process instrumentation activities. 
Each time the hostile process accesses the memory, the tool switching, above described, occurs again. 

Currently, we are trying to improve this technique in order to disarm \textit{debug-aware} malwares which once detect they are into a honeypot try to evade the system, for example by killing themselves.

%\texttt{{\char'134}section} that

%\begin{table}
%\centering
%\caption{Frequency of Special Characters}
%\begin{tabular}{|c|c|l|} \hline
%Non-English or Math&Frequency&Comments\\ \hline
%\O & 1 in 1,000& For Swedish names\\ \hline
%$\pi$ & 1 in 5& Common in math\\ \hline
%\$ & 4 in 5 & Used in business\\ \hline
%$\Psi^2_1$ & 1 in 40,000& Unexplained usage\\
%\hline\end{tabular}
%\end{table}

\section{Experiments}
\label{sec:exp}
\textsc{HERESy} has been tested by distributing it inside a \texttt{/16} network (i.e. a campus network). Thousands of attacks targeting various protocols have been detected. Among them, the most targeted were FTP, SSH, TELNET, HTTP and SMB.
%
%
 %We were able to relief if the behavior was human or automated. Once a connection starts, the proxy module builds up an isolated environment on the fly and all the attacker’s traffic is redirected to it.Then, all the network traffic gets dumped by the network module per environment basis. 
 % 
 Independently from the kind of attack, the system is able to recognize and log all the spawned commands. If the executed binaries belong to the attacked environment, their execution is just logged, otherwise, i.e. the process is recognized alien, the dynamic binary instrumentation module begins tracing it. All the instructions executed are saved in a separate log for each traced process. Furthermore, an experimental process dumper takes periodic snapshots saving all the information regarding file descriptors, memory, registers, and so on. All the file system accesses, which concern modifications, are saved on a versioning basis. All this workflow allowed to observe almost completely the behavior of various malwares and  permitted to accelerate the reverse engineering process.
 %having something really near to the dynamic malware analysis. 
 The vast majority of the captured attacks  comes from the BRIC area, most of them targeting embedded systems, many others have as target common platforms like \texttt{x86\_64} and others were equipped of multi-platform droppers written in common script languages such as bash, Python, etc. The  deployment of \textsc{HERESy} instances  in various machines inside this big network permitted to observe that the behavior of some malware is also iterative, in the sense that an attack on a low IP address inside the same network reached high  IP number addresses just some minutes later. So, we observed basically that this kind of malware tries to reproduce itself enumerating many IP addresses and searching for the same vulnerability.

\subsection{A simple malware}

\begin{figure}[btp]
\includegraphics[width=1.0\columnwidth]{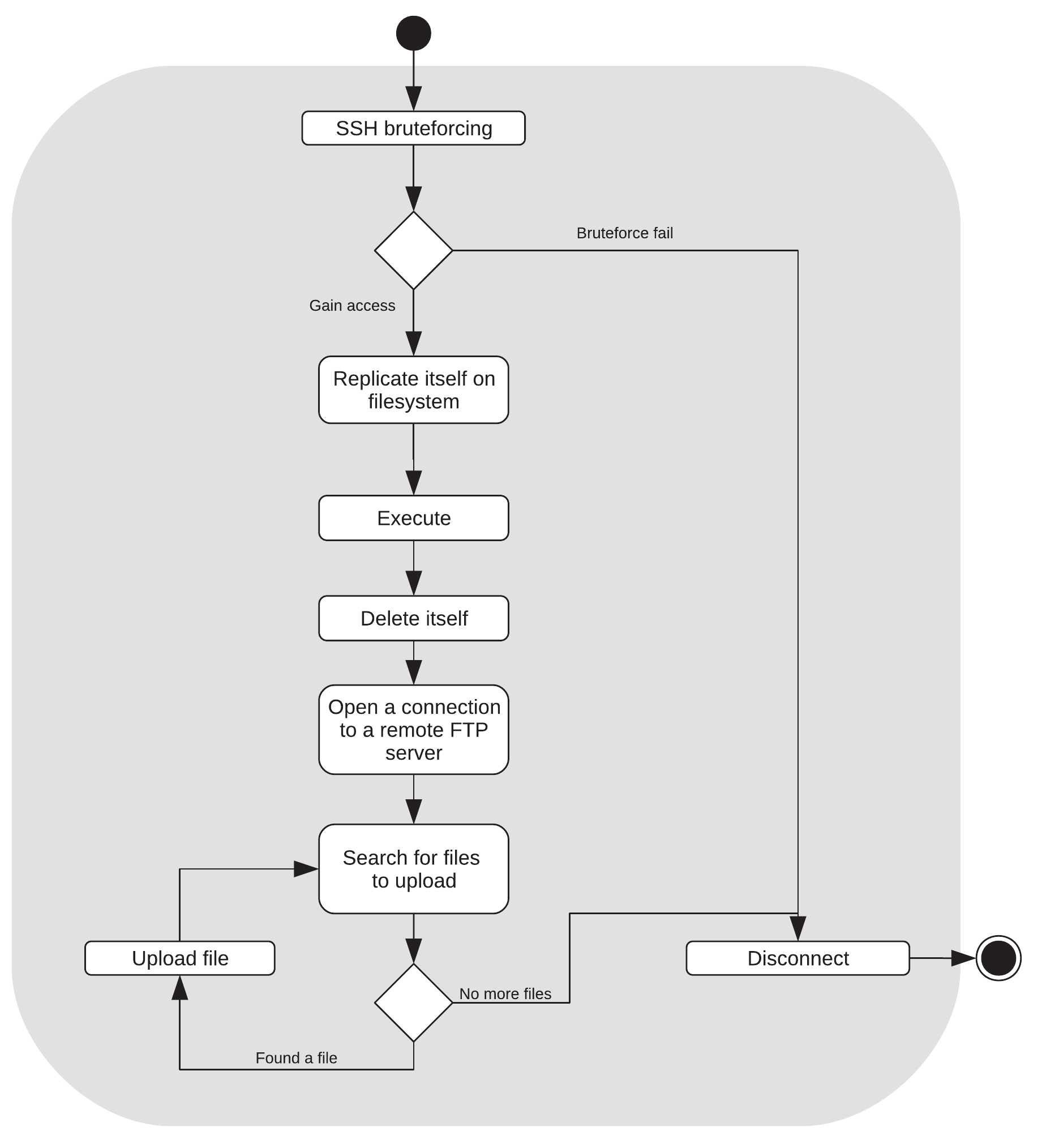}
\caption{Behavior of a captured malware}
\label{fig:malwb}
\end{figure}

During the testing phase of \textsc{HERESy}, we captured a series of malwares. In this section, we will describe the most simple of them.
We analyzed the his behavior by extracting the data captured by \textsc{HERESy}. The malware entered the honeypot through a brute force attack to the SSH protocol. Once it guessed the password, it reproduced itself on a specific location of the file system. So we tracked the copy of the malicious binary inside our file system replica. Then, the network module sniffed all the connections to a malicious FTP server performed by the malware and controlled by an attacker. By looking at the memory dump of the malware, we found out a behavior that it was not observed  during the very first static analysis phase. The malware had various encrypted strings representing well known document extensions. Observing the memory into his various modifications states revealed those strings and looking at the dumped code accessing this memory block we observed the intention to search and upload the files matching such extensions. When the malware ends its execution it deletes itself and his copies from the files system, however the \textsc{HERESy} instrumentation  allowed us to save these binaries and the other replicas. The behavior of the captured malware is sketched in the activity diagram of Fig.~\ref{fig:malwb}

\subsection{Real Time  Monitoring}
To observe the behavior of \textsc{HERESy} we developed a centralized system to gain and transmit information about  its operation. Each module (except the networking module) sends updates about new connections, command executed and file modifications to a daemon which in turns acts as a web server. Our clients connects to this daemon through \texttt{WebSockets}~\cite{websock2013} and monitors in real time all the actions performed by attackers/malwares inside the isolated environments. Using this strategy we can observe the overall system status and also it come useful to debug its behavior. So we deploy multiple instances of the \textsc{HERESy} architecture inside our huge network using our cloud or machine which are up for the purpose and we are constantly aware of the threats which menace our institutional infrastructures.
The information generated by the  modules is also stored  into a NoSQL database~\cite{nosql} in order to allow to query off-line the system and make statistics about the attacks received every day. The employed database is ElasticSearch~\cite{gormley2015} because of its good performances. We observed that it is able to execute a thousand of writes per seconds without a significant impact on performance of the machine used for storage purposes.

%\cite{Hazelwood2011}

\section{Conclusions}
\label{sec:conc}

In this paper we presented \textsc{HERESy}, a honeypot software system born to support and enhance the effectiveness of malware reverse engineering activities. \textsc{HERESy} is able to create on-demand high-interaction isolated environment which are exploited to capture  malware behavior and store the byproducts of its activity. \textsc{HERESy} features have proven to be useful tools for easing and powering malware analysis tasks.

Future improvements, which are already in experimentation, are adding support for process memory dump at specific points of execution and the possibility to navigate through these snapshots in order to restore a particular status of the malicious process inside the environment. Another characteristic on which we are currently working on is a procedure to automatically generate signatures for new malwares and communicate these to IDS in order for them to automatically recognize and block future attacks. During the \textsc{HERESy} development and the subsequent testing phase we captured a huge number of malwares and we studied their behavior using the above cited instruments. The main features of \textsc{HERESy}, i.e. the ability of dumping file system modifications and the network and process tracking, greatly relieved our work on threat analysis.

%ACKNOWLEDGMENTS are optional
%\section{Acknowledgments}
%\input{ack.tex}

\bibliographystyle{abbrv}
\bibliography{references}  % sigproc.bib is the name of the Bibliography in this case
% ACM needs 'a single self-contained file'!
%\balancecolumns

\end{document}